\documentclass[aps,prd,twocolumn,superscriptaddress,nofootinbib,showpacs,10pt]{revtex4-1}
\usepackage[left=2.0cm,right=2.0cm,top=2.0cm,bottom=2.0cm,marginparwidth=17mm]{geometry}
\usepackage[utf8]{inputenc}
\usepackage[english]{babel}
\usepackage{amsmath}
\usepackage{bbm}
\usepackage{amsfonts}
\usepackage{amssymb}
\usepackage{graphics}
\usepackage{graphicx}
\usepackage{subfigure}
\usepackage[usenames,dvipsnames,svgnames]{xcolor}  
\usepackage[linktocpage]{hyperref}   
\usepackage{todonotes}
\usepackage{ulem} 
\usepackage{braket}
\hypersetup{colorlinks=true,linkcolor=beamer@PRD, citecolor=beamer@PRD}

\renewcommand\eqref[1]{\textcolor{beamer@PRD}{(}\ref{#1}\textcolor{beamer@PRD}{)}}

\definecolor{beamer@PRD}{RGB}{46,48,146}
\begin{document}
\title{Improved coherence time of a non-Hermitian qubit in a $\mathcal{PT}$-symmetric Environment}
\author{Duttatreya}
\affiliation{Department of Physics, BITS-Pilani, K K Birla Goa Campus, Zuarinagar, Goa 403726, India}
\author{Ipsika Mohanty}
\affiliation{Institute for Theoretical Solid State Physics, RWTH Aachen University, Aachen 52074, Germany}
\author{Sanjib Dey}\email{sanjibd@goa.bits-pilani.ac.in}
\affiliation{Department of Physics, BITS-Pilani, K K Birla Goa Campus, Zuarinagar, Goa 403726, India}
\begin{abstract}
Quantum computing's potential for exponential speedup is fundamentally limited by decoherence, a phenomenon arising from environmental interactions. Non-Hermitian quantum mechanics, particularly $PT$-symmetric systems, offers a novel framework for extending coherence times. This study examines a qubit's coherence under non-Hermitian $PT$-symmetric dynamics, highlighting significantly enhanced coherence times compared to Hermitian setups. The effect is especially pronounced when both the system and environment exhibit $PT$-symmetry. Interestingly, greater environmental non-Hermiticity correlates with extended coherence, contrary to traditional expectations. These findings point to promising strategies for managing decoherence, which could significantly advance approaches to quantum information processing.
\end{abstract}


\maketitle
\section{Introduction} \label{sec1}
Quantum computing's promise of exponential computational speedup relies fundamentally on the ability to maintain quantum coherence. The coherence of a qubit is characterized by two primary timescales: the energy relaxation time $T_1$, which describes the decay from excited to ground state, and the dephasing time $T_2$, which characterizes the loss of phase relationships between quantum states. The overall coherence time $T_2^*$ combines both pure dephasing and energy relaxation effects, following the relationship $\frac{1}{T_2^*} = \frac{1}{2 T_1} + \frac{1}{T_2}$ \cite{Schlosshauer_2007_Book,chirolli_burkard_2008}. In theoretical studies of decoherence, pure dephasing ($T_2$) often takes precedence as it typically occurs more rapidly than energy relaxation and represents the fundamental limitation on quantum information preservation. Moreover, pure dephasing provides a cleaner theoretical framework for studying environmental interactions, as it isolates the phase destruction mechanism from energy exchange processes \cite{Breuer_Petruccione_2002_Book}.

The interaction between quantum systems and their environment, which leads to decoherence, is formally studied within the framework of open quantum systems. This approach recognizes that perfect isolation of quantum systems is impossible, necessitating methods to describe their coupling to the environment. The most common theoretical tools for modeling open quantum systems include master equations in Lindblad form \cite{Gorini_Kossakowski_Sudarshan_1976, Lindblad_1976}, quantum trajectories \cite{Belavkin_Barchielli_1991, Plenio_Knight_1998}, and influence functional methods \cite{Feynman_Vernon_2000}. These approaches typically treat the environment as a collection of harmonic oscillators or spins, with the system-environment interaction leading to irreversible information loss. Traditional studies of decoherence through these methods have shown that environmental coupling generally accelerates the loss of quantum coherence \cite{Plenio_Knight_1997}.

Compelling ideas for slowing down the decoherence have been proposed with different quantum mechanical \cite{Vitali_Tombesi_Milburn_1997,Horoshko_Kilin_1997,Viola_Lloyd_1998, Thorwart_Hartmann_Goychuk_Hanggi_2000,Agarwal_Scully_Walther_2001,Branderhorst_etal_2008, Poletti_Bernier_Georges_Kollath_2012}, $PT$-symmetric \cite{Gardas_Deffner_Saxena_2016, Fring_Frith_2019, Chakraborty_Sarma_2019} and deformed \cite{Dehdashti_etal_2015} setups. Non-Hermitian quantum mechanics has emerged as a powerful alternative framework for describing open quantum systems, offering insights beyond conventional Hermitian approaches \cite{Gardas_Deffner_Saxena_2016, Dey_Raj_Goyal_2019}. A particularly intriguing subset of non-Hermitian systems are those possessing $PT$-symmetry, where $P$ represents parity reversal and $T$ denotes time reversal. $PT$-symmetric systems can exhibit entirely real energy spectra despite their non-Hermitian nature \cite{Bender_Boettcher_1998}, a property that has profound implications for quantum dynamics. The $PT$-symmetric framework naturally accommodates the description of balanced gain and loss, where energy-increasing (gain) and energy-decreasing (loss) processes occur symmetrically within the system. This gain-loss formalism has found successful applications in optical systems \cite{Guo_etal_2009, Ruter_etal_2010}, where it has led to the observation of exceptional points and novel wave propagation effects. In quantum systems, $PT$-symmetry offers a unique perspective on environmental interactions, suggesting the possibility of manipulating decoherence through carefully engineered gain-loss arrangements \cite{Gardas_Deffner_Saxena_2016}. Numerous theoretical studies \cite{Longhi_2009, Graefe_Jones_2011, Dey_Fring_Gouba_2015} and various experiments found the existence of non-Hermiticity in several physical systems, most notably in photonics \cite{Feng_etal_2013}, quantum walk \cite{Xiao_etal_2017, Mittal_Raj_Dey_Goyal_2021}, acoustics \cite{Fleury_Sounas_Alu_2015}, microresonator \cite{Chang_etal_2014}, material science \cite{Weimann_etal_2017}, etc.

In this work, we demonstrate that a qubit exhibiting non-Hermitian $PT$-symmetric evolution, when coupled to a non-Hermitian $PT$-symmetric environment, displays remarkably extended coherence times compared to conventional Hermitian configurations. Our results show that this enhancement persists even when only one component (either the qubit or the environment) maintains $PT$-symmetry, though the effect is most pronounced when both components are $PT$-symmetric. Surprisingly, we find that increasing the non-Hermiticity of the environment leads to longer coherence times, contradicting the conventional wisdom that stronger environmental coupling accelerates decoherence. These findings suggest new strategies for protecting quantum information through engineered non-Hermitian environments and shed light on the fundamental relationship between $PT$-symmetry and quantum coherence.

The paper is organized as follows: In Sec.\,\ref{sec2}, we establish the theoretical framework of Kraus operators for quantum interacting systems with pure dephasing. Sec.\,\ref{sec3} introduces our non-Hermitian interacting model, detailing the mathematical structure of both the system and environment Hamiltonians and their coupling. The evolution dynamics are calculated in Sec.\,\ref{sec4}, where we derive the key expression for the coherence time. Sec.\,\ref{sec5} presents an in-depth scrutiny of the decoherence process, examining the role of $PT$-symmetry and environmental non-Hermiticity in extending coherence times. In Sec.\,\ref{sec6}, we propose an experimental schematic for implementations of our theoretical framework, discussing practical considerations and potential platforms for observing these effects. Our work concludes with a discussion of implications for quantum computing and future research directions.

\section{interacting systems with pure dephasing}\label{sec2}
Interacting quantum models are composed of two quantum bodies, a large bath/environment $E$ and a small system $S$ (enclosed within the environment $E$), whose dynamics are governed by the free Hamiltonians $\mathcal{H}_E$ and $\mathcal{H}_S$, respectively. The individual sub-bodies $S$ and $E$ live in the Hilbert spaces $\mathbb{H}_S$ and $\mathbb{H}_E$, correspondingly, and, hence, the composite body $S+E$ resides in the combined higher dimensional Hilbert space $\mathbb{H}_S \otimes \mathbb{H}_E$. The combined body $S+E$ may be recognized as an isolated entity evolving unitarily; however, the sub-bodies $S$ and $E$ are involved in mutual interaction leading to non-unitary dynamics for each of them. The system-environment interaction can be described by an additional Hamiltonian $\mathcal{H}_{\text{int}}$ so that the total Hamiltonian of the interacting model turns out to be $\mathcal{H}=\mathcal{H}_S+\mathcal{H}_E+\mathcal{H}_{\text{int}}$.

Assume that, at the beginning, the environment is in the vacuum state and the system and environment are uncorrelated such that the initial state of the combined body $S+E$ can be represented by a separable density matrix $\rho_{SE}(0)=\rho_{S}(0) \otimes \rho_{E}(0)=\rho_{S}(0) \otimes |0\rangle_E{}_E\langle 0|$. So, its evolution is given by $\rho_{SE}(t)=\mathcal{U}_{SE}(t)\rho_{SE}(0)\mathcal{U}_{SE}^\dagger (t)$, with $\mathcal{U}_{SE}(t)$ being a unitary time propagator. The reduced system dynamics, which is of our main interest, can be obtained by eliminating the environment degrees of freedom from the total density matrix as
\begin{equation}
\rho_{S}(t)=\mathrm{Tr}_{E}\{\rho_{SE}(t)\}=\sum_i \mathcal{K}_i \rho_S(0) \mathcal{K}_i^\dagger,
\end{equation}
where $\mathcal{K}_i=\langle i|\mathcal{U}_{SE}(t)| 0\rangle_E$ is the Kraus operator with $\{|i\rangle\}$ being an orthonormal basis of $\mathbb{H}_E$. The Kraus operators, therefore, act only on $\mathbb{H}_S$, and their matrix representation is realized as $\mathcal{K}_i^{mn}=\langle m|\mathcal{K}_i|n\rangle$, where $|m\rangle$ and $|n\rangle$ are defined on $\mathbb{H}_S$. Phase damping is a convenient mechanism that comprehends the description of system-environment interaction by first splitting the gigantic environment into a large number of tiny sub-environments and then assembling all the system-sub-environment interactions back together. Naturally, within this approach, the interaction between the system and the environmental subsystems can be regarded much weaker compared to the system energy but stronger than that of the sub-environment. Thus, for a unitary evolution, the interaction effectively leaves the system unchanged but modifies each of the sub-environments and, hence, the environment. As for example, consider that the photon (system) and the dust particle (environment) initially are in the states $|0\rangle_S$ and $|0\rangle_E$, respectively. After the evolution, there is a probability $p$ that the state of the dust particle moves to the first excited state $|1\rangle_E$, which means $\mathcal{U}_{SE}(t)|0\rangle_S|0\rangle_E\rightarrow\sqrt{1-p}|0\rangle_S|0\rangle_E+\sqrt{p}|0\rangle_S|1\rangle_E$. Whereas, if initially, the system would have been in the state $|1\rangle_S$, and the environment in $|0\rangle_E$ then $\mathcal{U}_{SE}(t)|1\rangle_S|0\rangle_E\rightarrow\sqrt{1-p}|1\rangle_S|0\rangle_E+\sqrt{p}|1\rangle_S|2\rangle_E$. Consequently, the Kraus operators become
\begin{equation}
\mathcal{K}_0=\sqrt{1-p}~\mathbbm{1},~
\mathcal{K}_1=\begin{pmatrix} 
\sqrt{p} & 0 \\
0 & 0 
\end{pmatrix},~
\mathcal{K}_2=\begin{pmatrix} 
0 & 0 \\
0 & \sqrt{p} 
\end{pmatrix},
\end{equation}
and, therefore, the density operator of the reduced system turns out to be
\begin{equation}\label{Density2}
\rho_S(t)=\sum_{i=0}^{2}\mathcal{K}_i\rho_S(0)\mathcal{K}_i^\dagger=\begin{pmatrix} 
\rho_S^{00} & (1-p)\rho_S^{01} \\
(1-p)\rho_S^{10} & \rho_S^{11} 
\end{pmatrix}.
\end{equation}
Here, we have introduced the matrix elements $\rho_S^{ij}=\langle i|\rho_S(0)|j\rangle$. Suppose that the evolution is continuous in time and $\Lambda$ is the probability of the environmental modification per unit time. Then, $p=\Lambda\Delta t$ is the probability of one such event during the time $\Delta t$, and we have $n$ such events in time $t$. The off-diagonal terms of \eqref{Density2}, then, turn out to be proportional to $(1-p)^n=(1-\Lambda\Delta t)^{t/\Delta t}\approx e^{-\Lambda t}$ resulting in
\begin{equation}\label{DensityMatrix}
\rho_S(t)=\begin{pmatrix} 
\rho_S^{00} & e^{-\Lambda t}\rho_S^{01} \\
e^{-\Lambda t}\rho_S^{10} & \rho_S^{11} 
\end{pmatrix}.
\end{equation}
In this method, the system population/energy is conserved since the diagonal elements of \eqref{DensityMatrix} are time-independent. On the other hand, the off-diagonal elements being decayed over time, the phase coherence of the system is lost; hence, the name pure dephasing. The loss of phase coherence (i.e. the decoherence) is governed by $\Lambda$ and, therefore, $\Lambda$ is called the decoherence factor.

Two kinds of open quantum systems are widely studied \cite{Breuer_Petruccione_2002_Book}, one in which a system is coupled to a bosonic environment being familiar as the spin-boson model, and the other in which environment is modeled in terms of spins, also known as the spin-environment model. The spin-boson model is valid in the high-temperature limit, whereas the spin environment is valid in ultra-low temperatures.  In this paper, we focus on the spin-boson model. The spin-boson model consists of a two-level spin system coupled to a large reservoir of bosonic modes that are modeled by large number of non-interacting harmonic oscillators. The free Hamiltonian of the system and that of the environment in thermal equilibrium are given by $\mathcal{H}_{S}=\omega_0\sigma_z/2$ and $\mathcal{H}_E=\sum_{i}\omega_{i}a_{i}^\dagger a_{i}$, respectively, where $\sigma_z$ denotes the Pauli matrix along the $z$-axis and, $\omega_0,~\omega_{i}$ are the natural frequencies of the corresponding entities. The system and environment are coupled via a bilinear interaction given by $\mathcal{H}_{\text{int}}=\sigma_{z} \otimes \sum_{i}(c_{i}a_{i}^{\dagger}+c_{i}^*a_{i})$, with $c_{i}$ being a complex parameter that represents the coupling strength between the system and the environment. Here, $a_{i}$ and $a_{i}^{\dagger}$ represent the bosonic ladder operators for the $i^{\text{th}}$ mode of the field. We have addressed all the terms in natural units $(\hbar =1)$.
\section{Non-Hermitian interacting model}\label{sec3}
A certain class of non-Hermitian Hamiltonians exhibit real eigenspectra and constitute well-defined physical systems \cite{Mostafazadeh_2002}. The necessary and sufficient condition for a non-Hermitian Hamiltonian $\mathcal{H}^{\text{NH}}$ to exhibit real eigenvalues is the existence of a non-unique positive semi-definite  metric operator $G=\eta^{\dagger}\eta$ satisfying $\mathcal{H}^{\text{NH}\dagger} G=G \mathcal{H}^{\text{NH}}$. It has also been shown that for any such non-Hermitian Hamiltonian, the matrix $\eta$, maps the non-Hermitian Hamiltonian $\mathcal{H}^{\text{NH}}$ onto a Hermitian Hamiltonian $\mathcal{H}$ through a similarity transformation $\mathcal{H}=\eta \mathcal{H}^{\text{NH}} \eta^{-1}$. The eigenspectrum remains invariant under such similarity transformation. The eigenstates of Hermitian Hamiltonian $\ket{\phi}$ are mapped to those of the non-Hermitian Hamiltonian $\ket{\phi^{\text{NH}}}$, using the same similarity transformation, as $\ket{\phi}=\eta \ket{\phi^{\text{NH}}}$. For a system coupled to an environment, the metric operator for the composite system is given by the similarity transformation $\eta=\eta_{S} \otimes \eta_{E}$, where $\eta_{S}$ and $\eta_{E}$ are the similarity transformations corresponding to the system and environment Hamiltonians, respectively. 

A non-Hermitian spin-boson model can be constructed by coupling a non-Hermitian two-level spin system with a non-Hermitian bosonic environment. An experimentally relevant non-Hermitian qubit system motivated by \cite{Gao_etal_2015} is given by
\begin{equation}\label{NHSystem}
\mathcal{H}_S^{\text{NH}} = \alpha \sigma_+ \sigma_- + \gamma \sigma_+ + \alpha^{*} \sigma_- \sigma_+ \gamma^{*} \sigma_-,
\end{equation}
where $\alpha,~\gamma$ are complex and $\sigma_+,~\sigma_-$ are fermionic raising and lowering operators, respectively. Without any loss of generality, $\alpha_S$ can be considered to be purely imaginary, renamed to $i \alpha_S$, and $\gamma=1$. Under this setting, $\mathcal{H}_S^{\text{NH}}=\sigma_x+i \alpha_S \sigma_z$ with $\sigma_x$ and $\sigma_z$ being the Pauli-$x$ and Pauli-$z$ matrices. The eigenvalues of this non-Hermitian Hamiltonian are given by $E_{1,2}=\pm \sqrt{1-\alpha_S^2}$, which is real as long as $|\alpha_S| \leq 1$. Consider a similarity transformation matrix of the form $\eta_S=e^{\vartheta \sigma_y}$, where the parameter $\vartheta$ is related to the non-Hermiticity parameter $\alpha_S$ as $\vartheta=\frac{1}{2}\tanh^{-1}(\alpha_S)$. The corresponding Hermitian Hamiltonian for the system can be obtained as
\begin{equation}\label{SystemHermitian}
\mathcal{H}_S=\eta_S \mathcal{H}_S^{\text{NH}} \eta_S^{-1}= E_1 \sigma_x.
\end{equation}
Let us introduce the non-Hermitian environment
\begin{equation}\label{BathNonHermitian}
\mathcal{H}_E^{\text{NH}}=\sum_{i} \left\{(\zeta-4\delta\tau^2)\frac{p_{i}^2}{2 m}+\frac{\delta}{2}k x_{i}^2+ i \delta\tau \omega_{i} (x_{i} p_{i} +p_{i} x_{i})\right\},
\end{equation}
with $\omega_{i}=\sqrt{k/m}$, so that its Hermitian counterpart becomes
\begin{equation}\label{BathHermitian}
\mathcal{H}_E=\eta_E \mathcal{H}_E^{\text{NH}}\eta_E^{-1}=\sum_{i} \left(\frac{\zeta}{2m}p_{i}^2+\frac{\delta}{2}kx_{i}^2\right).
\end{equation}
where $\eta_E=e^{\tau \sum_{i} \frac{p_{i}^2}{m \omega_i}}$ with $\zeta,\delta$ being non-zero generic functions of the real parameter $\tau$. Note that the Hermitian counterpart of the environment \eqref{BathHermitian} coincides with the bosonic environment when $\zeta=\delta=1$ and, in this case, the equation \eqref{BathNonHermitian} becomes simpler. However, we have kept $\zeta$ and $\delta$ generic so that we can study their additional roles in controlling the decoherence, if any. The Hermitian and non-Hermitian counterparts and of the interaction Hamiltonian take the form $\mathcal{H}_{\text{int}}=E_1\sigma_{x} \otimes \sum_{i}(c_{i}a_{i}^{\dagger}+c_{i}^*a_{i})$ and  $\mathcal{H}_{\text{int}}^{\text{NH}}= (\sigma_x + i \alpha_S \sigma_z) \otimes \sum_{i} \{c_{i} a_{i}^\dagger +c_{i}^* a_{i} +\tau (c_{i}+c_{i}^*) (a_{i}-a_{i}^{\dagger})\}$, respectively, with respect to the combined transformation $\eta=e^{\vartheta \sigma_y} \otimes e^{\tau \sum_{i} \frac{p_{i}^2}{m \omega_i}}$. Therefore, the total Hamiltonian for the non-Hermitian spin-boson model is given as
\begin{eqnarray}
\mathcal{H}^{\text{NH}} &=& \sigma_x + i \alpha_S \sigma_z + 
\sum_{i} \omega_{i}\left\{\Theta_+ a_{i}^2 + \Theta_- a_{i}^{\dagger2}\right. \nonumber \\
&& \quad \left. + \left(\frac{\zeta+\delta}{4} - \delta \tau^2\right)(2 a_{i}^{\dagger}a_{i}+1) \right\} \nonumber \\
&& + (\sigma_x + i \alpha_S \sigma_z) \otimes \sum_{i} \left\{c_{i} a_{i}^\dagger +c_{i}^* a_{i} \right. \nonumber \\
&& \quad \left. + \tau (c_{i}+c_{i}^*) (a_{i}-a_{i}^{\dagger}) \right\},
\end{eqnarray}
where $\Theta_\pm=\delta(\tau\pm 1/2)^2-\zeta/4$. The corresponding Hermitian Hamiltonian is given as
\begin{eqnarray}\label{InteractionHermitian}
\mathcal{H}&=&E_1 \sigma_x + E_1 \sigma_{x} \otimes \sum_{i}(c_{i}a_{i}^{\dagger}+c_{i}^*a_{i})\\
&&+\sum_{i} \omega_{i}\left\{\frac{\zeta+\delta}{2}(a_{i}^{\dagger}a_{i}+\frac{1}{2})+\frac{\delta-\zeta}{4}  (a_{i}^2+a_{i}^{\dagger2})\right\}.\nonumber
\end{eqnarray}
\section{Evolution and Decoherence}\label{sec4}
The reduced system dynamics can be obtained by virtue of the interaction picture. We can use two-equivalent methods for this. We can either evolve the non-Hermitian counterpart of the Hamiltonian $\mathcal{H}_S^{\text{NH}}$, $\mathcal{H}_E^{\text{NH}}$, $\mathcal{H}_{\text{int}}^{\text{NH}}$ by reconstructing the Dyson mapping in biorthogonal basis. Alternatively, we can stick to the formulations of standard quantum mechanics and evolve the Hermitian counterparts of the Hamiltonians $\mathcal{H}_S$, $\mathcal{H}_E$ and $\mathcal{H}_{\text{int}}$. For the sake of convenience, we stick to the second approach so that we obtain
\begin{equation}
    \mathcal{H}_{\text{int}}(t)=e^{-i(\mathcal{H}_S+\mathcal{H}_E)t}\mathcal{H}_{\text{int}}e^{i(\mathcal{H}_S+\mathcal{H}_E)t},
\end{equation}
which when replaced with the Hamiltonians from \eqref{SystemHermitian}, \eqref{BathHermitian}, \eqref{InteractionHermitian}, and simplified, takes the from
\begin{eqnarray}
&&\mathcal{H}_{\text{int}}(t)=E_1 \sigma_x \otimes \sum_{i}\left[(c_{i} a_{i}^{\dagger}+c_{i}^* a_{i})\cos{(\Gamma_{i} t)}\right. \\
&&~~~~~ \left. +\left\{(\zeta+\delta) (c_{i} a_{i}^{\dagger}-c_{i}^* a_{i})+(\zeta-\delta) (c_{i} a_{i} - c_{i}^* a_{i}^{\dagger})\right\}\right.\nonumber \\ 
&&\qquad\qquad \left.\times\frac{i\omega_{i}}{2\Gamma_{i}}\sin(\Gamma_{i} t)\right],\nonumber
\end{eqnarray}
where $\Gamma_{i}=\sqrt{\zeta\delta}\omega_{i}$. Thereafter, the time-propagator can be procured from $\mathcal{U}_{SE}(t)= \mathcal{T}\exp\{-i\int_{0}^{t}dt' \mathcal{H}_{\text{int}}(t')\}$, where the symbol $\mathcal{T}$ represents the time ordering. A closed and exact form of $\mathcal{U}_{SE}$ is customarily attained after an expansion of the integral in the Dyson series followed by a simplification facilitated by the commutative property of the time-dependent interaction Hamiltonian $\mathcal{H}_\text{int}(t)$ at different times. Unfortunately, it is not always guaranteed that $[\mathcal{H}_\text{int}(t),\mathcal{H}_\text{int}(t')]=0$ and, indeed, this is the case with our model. However, thanks to $[a_{i},a_{i}^\dagger]=1$, which turns the commutators $[\mathcal{H}_\text{int}(t),\mathcal{H}_\text{int}(t')]$ into c-numbers
\begin{eqnarray}
&&[\mathcal{H}_{\text{int}}(t),\mathcal{H}_{\text{int}}(t')]=\sigma_x^2 E_1^2 \otimes \sum_{i}  \frac{\omega_{i}}{2\Gamma_{i}}\sin\{\Gamma_{i}(t'-t)\}\nonumber  \\
&&\qquad\qquad \times\left[(\zeta+\delta) |c_{i}|^2-(\delta-\zeta) (c_{i}^2+c_{i}^{*2})\right].
\end{eqnarray}
Therefore, the time-ordering $\mathcal{T}$ can be replaced by an overall time-dependent global phase as $\mathcal{U}_{SE}(t)=e^{i\varphi(t)}\exp[-i\int_{0}^{t}dt' \mathcal{H}_{\text{int}}(t')]$. A time-dependent global phase has no additional physical implication on the temporal dynamics of the system and, so, it can be ignored for the sake of simplicity. Thus, the time-propagator simplifies to
\begin{equation}
\mathcal{U}_{SE}(t)=\exp\left[\sigma_x \otimes \sum_{i} \left\{\mu_{i}(t) a_{i}^{\dagger}-\mu_{i}^*(t)a_{i}\right\}\right]
\end{equation}
with, 
\begin{eqnarray}\label{muk}
\mu_{i}(t)&=&\frac{E_1\omega_{i}}{\Gamma_{i}^2}\sin^2\left(\frac{\Gamma_{i} t}{2}\right)\left\{(\delta-\zeta) c_{i}^*-(\zeta+\delta) c_{i}\right\}\nonumber\\
&&\quad -\frac{i E_1 c_{i}}{\Gamma_{i}}\sin(\Gamma_{i} t).
\end{eqnarray}
We are now ready to work out the temporal dynamics of the reduced system with the aid of the matrix elements
\begin{equation}\label{MatrixElement}
\rho^{ij}_S(t)=\bra{i}\mathrm{Tr}_E\{\mathcal{U}_{SE}(t)\rho_{SE}(0) \mathcal{U}_{SE}^{\dagger}(t) \}\ket{j},
\end{equation}
which can be computed with the consideration that the steady state of the environment is the thermal state, viz. $\rho_E(0)=\prod_{i} \rho_{E,k}(T)=\prod_{i} \left(1-e^{-\beta\omega_{i}}\right)e^{-\beta\omega_{i}a_{i}^\dagger a_{i}}$, with $\beta$ denoting the thermodynamic beta/coldness. The off-diagonal elements of the reduced-system density matrix turn out to be 
\begin{equation}\label{OffDiagonal}
\rho_S^{01}(t)=\rho_S^{01}(0)\prod_{i}\left\langle\mathfrak{D}\left(\mu_{i}(t)\right)\right\rangle = \left[\rho_S^{10}(t)\right]^\ast.
\end{equation}
Here, we define $\mathfrak{D}(\mu_{i}(t))= e^{\mu_{i}(t)a_{i}^{\dagger}-\mu_{i}^*(t) a_{i}}$, whose expectation value can be identified as the Wigner characteristic function and can be simplified as
\begin{eqnarray}\label{Decoherence5}
\prod_{i}\left\langle\mathfrak{D}\left(\mu_{i}(t)\right)\right\rangle &=& \exp\left\{-\sum_{i} \frac{|\mu_{i}(t)|^2}{2}\coth\left(\frac{\omega_{i}}{2 K T}\right)\right\} \nonumber\\
&:=& e^{-\Lambda(t)},
\end{eqnarray}
where $K,\Lambda$ represent the Boltzmann constant and the decoherence factor, respectively. Note that the free Hamiltonian of the system $\mathcal{H}_S$ commutes with that of the interacting Hamiltonian $\mathcal{H}_\text{int}$ and, therefore, the environment does not have a role in modifying the population. It implies that the diagonal element of \eqref{MatrixElement} remain unaffected by the evolution, $\rho_S^{ii}(t)=\rho_S^{ii}(0)$. However, the environmental interaction essentially modifies the off-diagonal terms, which are calculated explicitly in \eqref{OffDiagonal}. It is straightforward to calculate the diagonal elements explicitly using the same method as those of the off-diagonal ones and verify that they remain unchanged after the evolution. Nevertheless, collecting $\mu_{i}(t)$ from \eqref{muk}, we obtain an exact form of the decoherence factor with the use of \eqref{Decoherence5} as follows
\begin{eqnarray}
\Lambda(t) &=& \sum_{i} |c_{i}|^2 \frac{E_1^2}{\Gamma_{i}^4} \Bigg[ \left(\zeta^2 \mathrm{Re}[c_{i}]^2 + \delta^2 \mathrm{Im}[c_{i}]^2\right) \nonumber \\
&\times&  4 \omega_{i}^2 \sin^4\left(\frac{\Gamma_{i} t}{2}\right) + \Gamma_{i}^2 \sin^2(\Gamma_{i} t) \nonumber \\
&+& 4 \Gamma_{i} \omega_{i}(\delta - \zeta)\mathrm{Re}[c_{i}]\mathrm{Im}[c_{i}]  \sin^2\left(\frac{\Gamma_{i}}{2} t\right)  \sin(\Gamma_{i} t)  \Bigg] \nonumber \\
&& \times  \coth\left(\frac{\omega_{i}}{2 K T}\right).
\end{eqnarray}
\section{Decoherence: Analysis in-depth}\label{sec5}
In this section, we provide a detailed study on the behavior of decoherence depending on the degrees of system and/or environment non-Hermiticity, initial state parameters and the strength of the interaction. Since the size of the environment is fairly large, the corresponding density modes can unambiguously be considered to be a continuous spectral density function $\mathcal{J}(\omega)$, i.e.
\begin{figure*}
  \subfigure[]{
\includegraphics[width=5.58cm]{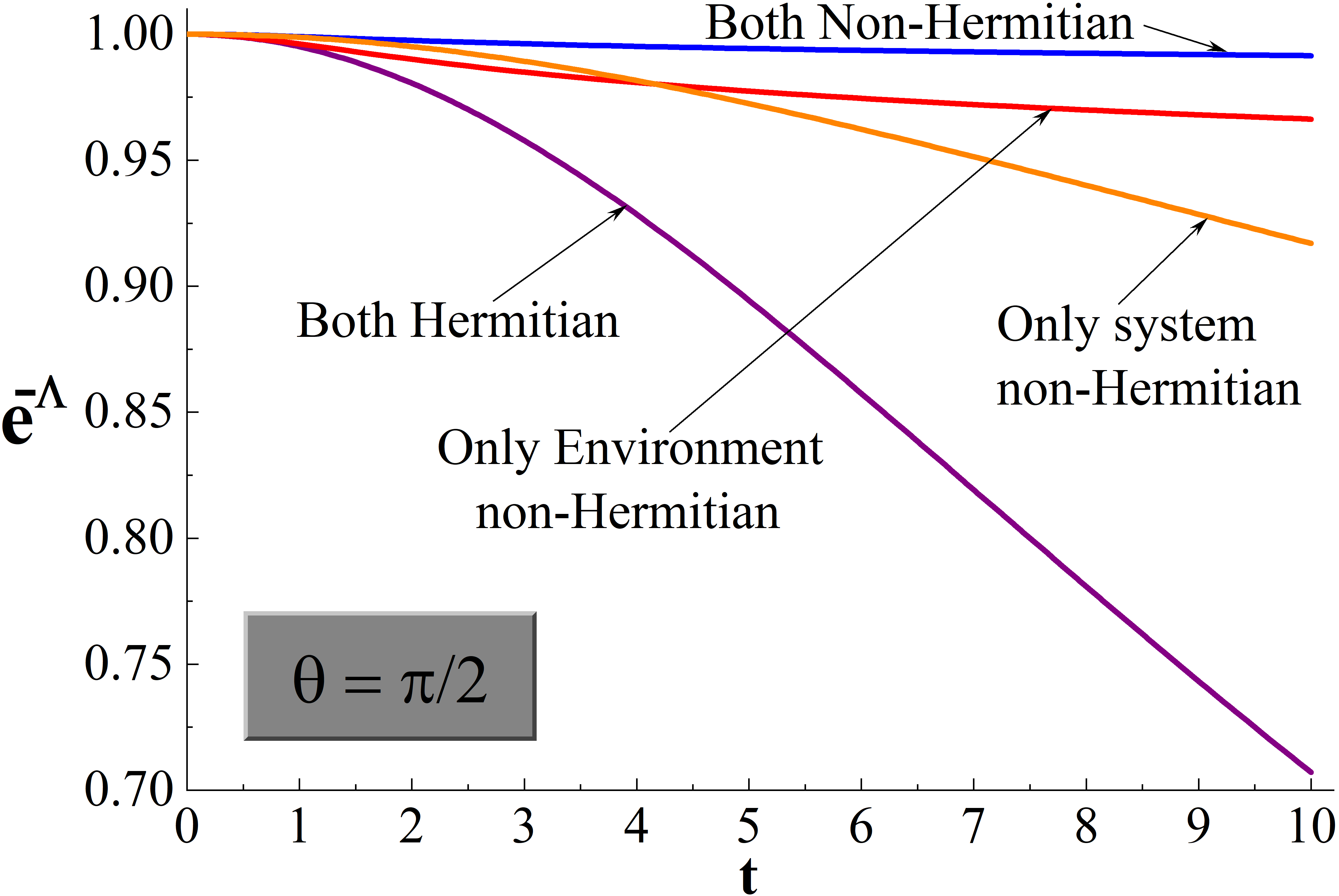}
\label{fig1a}}
\subfigure[]{
  \includegraphics[width=5.58cm]{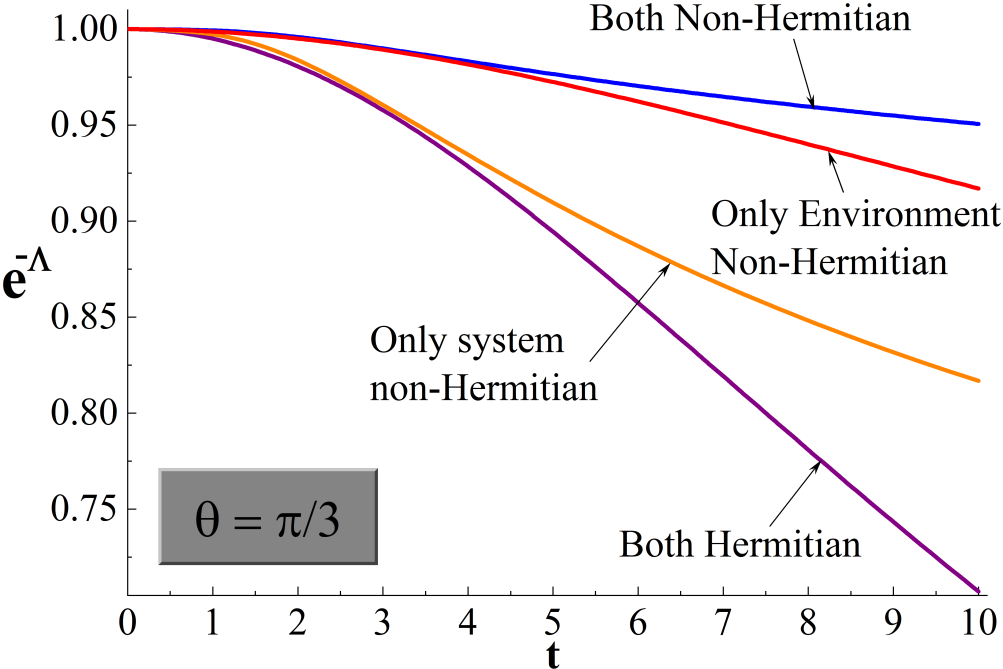}
  \label{fig1b}}
  \subfigure[]{
    \includegraphics[width=5.58cm]{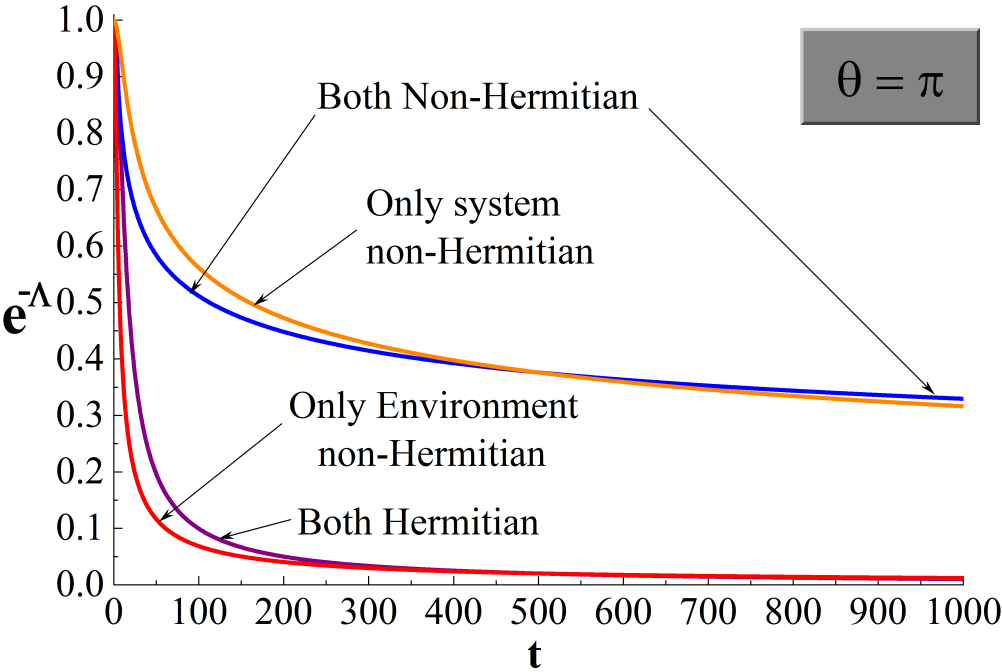}
    \label{fig1c}} 
 \caption{(Color online) Temporal variation of decoherence for Hermitian system and environment ($E_1=1, \tau=0$), non-Hermitian system and Hermitian environment ($E_1=0.5, \tau=0$), Hermitian system and non-Hermitian environment ($E_1=1, \tau=2$) and both system and environment non-Hermitian ($E_1=0.5, \tau=2$) for a fixed value of $\theta$ \subref{fig1a} $\theta=\pi/2$ \subref{fig1b} $\theta=\pi/3$ and \subref{fig1c} $\theta=\pi$.}
  \label{fig1}
\end{figure*}
\begin{equation}
\sum_{i} |c_{i}|^2\longrightarrow \int_{0}^{\infty} \mathcal{J}(\omega) d\omega.
\end{equation}
To convert the decoherence factor $\Lambda$ entirely in terms of the continuous spectral density function $\mathcal{J}(\omega)$, we express the coupling constant for each mode $c_{i}$ into the polar coordinates such that $\mathrm{Re}[c_{i}]^2 \rightarrow |c_{i}|^2 \cos^2\theta$, $\mathrm{Im}[c_{i}]^2 \rightarrow |c_{i}|^2 \sin^2\theta$ and $\mathrm{Re}[c_{i}]\mathrm{Im}[c_{i}] \rightarrow |c_{i}|^2 \sin\theta\cos\theta$. We also assume the density function $\mathcal{J}(\omega)$ to have a linear variation in the small frequency limit as well as an exponential cutoff in the high-frequency limit, i.e. $\mathcal{J}(\omega)=\mathcal{A}\omega e^{-\omega/\Omega}$. Here $\mathcal{A}$ is the proportionality constant and $\Omega$ is the measure of the high frequency cutoff. Thus, in the continuum frequency limit, the decoherence factor can be expressed as
\begin{eqnarray}\label{Decoherence_factor}
        \Lambda(t)&=&\int_{0}^\infty \frac{2 E_1^2}{ \Gamma^4}\mathcal{A}\omega e^{-\omega/\Omega} \Bigg[ \left(\zeta^2 \cos^2\theta + \delta^2 \sin^2\theta\right) \nonumber \\
        &\times&  4 \omega^2 \sin^4\left(\frac{\Gamma t}{2}\right) + \Gamma^2 \sin^2(\Gamma t) \nonumber \\
        &+& 4\Gamma \omega(\delta - \zeta)\cos\theta\sin\theta  \sin^2\left(\frac{\Gamma}{2} t\right)  \sin(\Gamma t) \Bigg] \nonumber \\
        &\times&  \coth\left(\frac{\omega}{2 K T} \right) d\omega. 
\end{eqnarray}
The decoherence factor for the standard Hermitian spin-boson system is easily obtained in the limit $\zeta=1$, $\delta=1$ and $E_1=1$ as \cite{Dey_Raj_Goyal_2019}
\begin{equation}
\lambda(t)=\int_{0}^{\infty} \frac{4 \mathcal{A} e^{-\omega/\Omega}}{\omega}\left\{1-\cos(\omega t)\right\}\coth\left(\frac{\omega}{2 K T}\right) d\omega.
\end{equation}
For the standard spin-boson model, it is already known that if we study the system dynamics with a pure state, the system eventually loses coherence with time. It has also been shown earlier \cite{Dey_Raj_Goyal_2019} that coupling a $\mathcal{PT}$-symmetric system to a Hermitian bosonic environment leads to slowing down of the decoherence process and similarly coupling a Hermitian topological system to a non-Hermitian bosonic environment also leads to similar effect \cite{Naghiloo_Abbasi_Joglekar_Murch_2019}. By numerically looking at the time dependence of the decoherence factor, we observe that there is a further slowing down of the decoherence process when a $\mathcal{PT}$-symmetric system is coupled to a non-Hermitian environment, as depicted in Fig.\,\ref{fig1}. 
\begin{figure}
\centering 
\includegraphics[width=7.2cm,height=4.7cm]{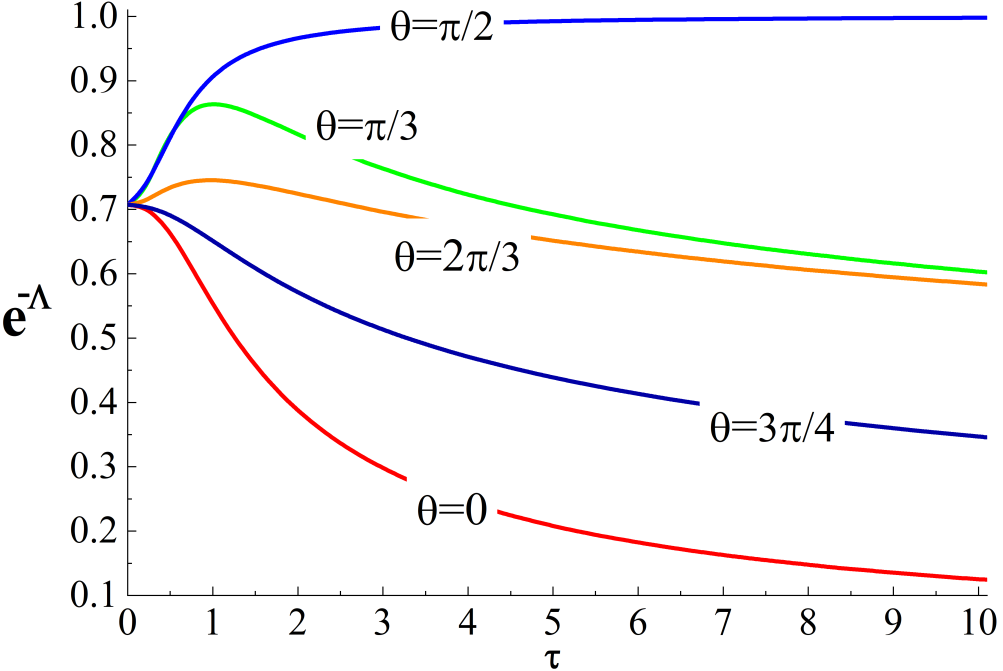}
\caption{(Color online) Variation of decoherence with non-Hermiticity of the environment for different values of $\theta$ for $E_1=1,~ t=10$.}
\label{fig5}
\end{figure}

In all three subfigures of the Fig.\,\ref{fig1}, we have demonstrated the temporal decoherence behavior in all possible distinct cases, namely, when both the system and environment are Hermitian, only system is non-Hermitian, only environment is non-Hermitian and both the system and environment are non-Hermitian. In each of the sub-figures, we have kept the value of $\theta$ fixed. Here, we observe two important phenomena. First, irrespective of the value of $\theta$, one obtains the best decoherence profile for the case when both the system and environment are non-Hermitian (as indicated by the blue line in all three sub-figures). Though, in Fig.\,\ref{fig1b} and \ref{fig1c}, we notice an intermediate faster decay of coherence for the shorter time of evolution; however, if we wait for a longer time for the evolution to be saturated, the coherence is maximally retained in the case when both the system and the environment are non-Hermitian. We have concluded it after testing all these plots for a significantly high value of time, but for convenience, we have presented them only up to the time beyond which the coherence never decays below the other three cases. The other important observation is that we obtain the maximal efficiency in preserving the coherence when the system and environment interacts at $\theta =\pi/2$. The same phenomenon is observed in Fig.\,\ref{fig5} too, where we have explicitly shown the dependence of the decoherence on the environment non-Hermiticity ($\tau$) for different values of $\theta$. The case, $\theta=\pi/2$ (blue line), in Fig.\,\ref{fig5} indicates the best decoherence profile. 
Following this, we have kept $\theta=\pi/2$ in all our subsequent studies for good reasons. Also, for all the figures, we have used $\mathcal{A}=1,~\Omega=0.1$ and $T=300$, without any loss of generality. 
\begin{figure}
\centering 
\includegraphics[width=7.2cm,height=4.7cm]{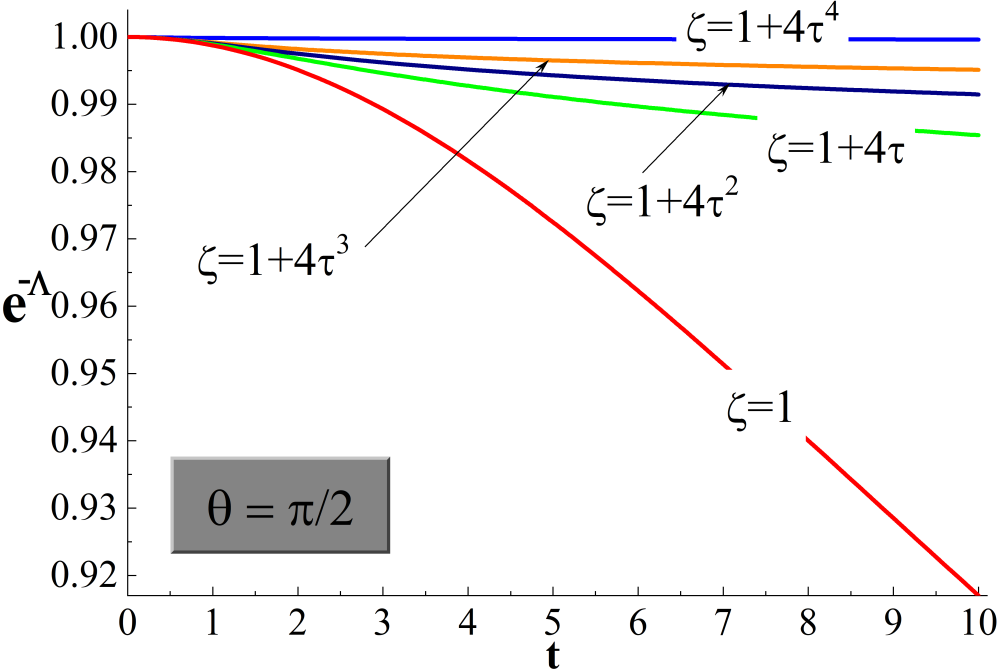}
\caption{(Color online) Variation of decoherence with time for different functional forms of environment non-Hermiticity $\zeta~\text{with}~\tau=2$  keeping the system non-Hermiticity fixed, i.e., $E_1=0.5$.}
\label{fig2}
\end{figure}

In obtaining the above figure and the rest of the figures, we have assume the functional dependence of $\zeta$ of $\tau$ to be of the particular form $\zeta=1+4 \tau^2$ and $\delta=1$. The reason is that this special choice of $\zeta$ and $\delta$ reduce the non-Hermitian environment Hamiltonian to the form $\mathcal{H}_e^{\text{NH}}=\sum_{i} \frac{p_{i}^2}{2 m}+ \frac{1}{2} k x^2+i \tau \omega (x p_{i} +p_{i} x)$ which is one of the well-studied Hamiltonians in the literature of non-Hermitian physics. However, from Fig.\,\ref{fig2}, we notice that for higher order dependence of $\zeta$ on $\tau$, the decoherence process slows down further.

\begin{figure}
\centering 
\includegraphics[width=7.2cm,height=4.7cm]{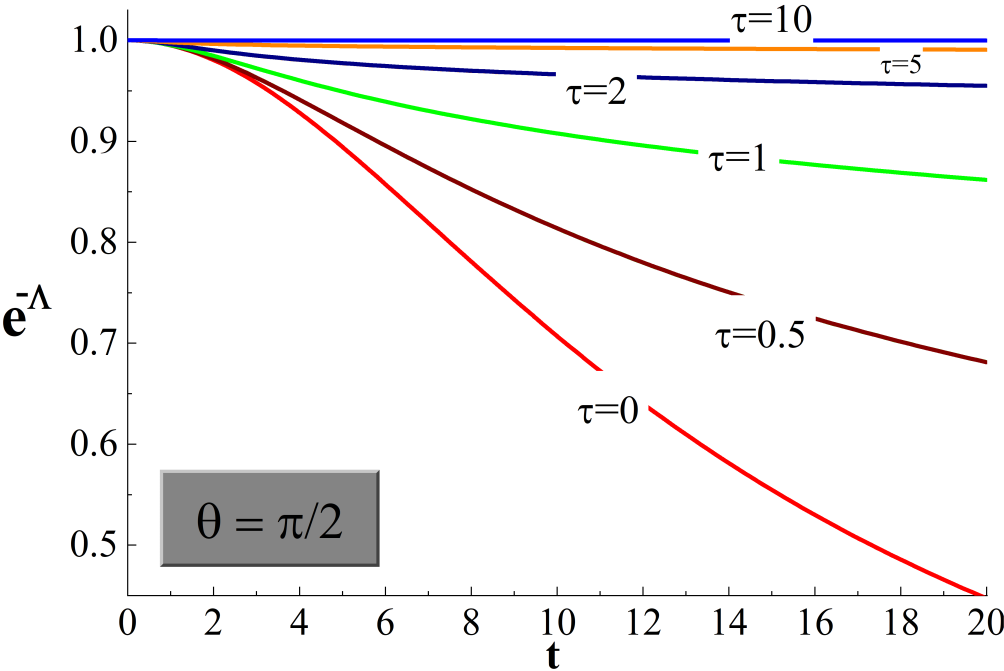}
\caption{(Color online) Time evolution of decoherence for different values of non-Hermiticity in the environment, keeping the Hermiticity of the system fixed, i.e., $E_1=1$.}
\label{fig4}
\end{figure}

The decoherence pattern in Fig.\,\ref{fig4} is somewhat obvious and it could be guessed from the pattern obtained in Fig.\,\ref{fig5}. However, for the sake of completeness, we have demonstrated it. Here, we notice that when the environment non-Hermiticity is significantly large for a fixed system Hermiticity, the coherence is preserved.  

\begin{figure}
\centering 
\includegraphics[width=7.2cm,height=4.7cm]{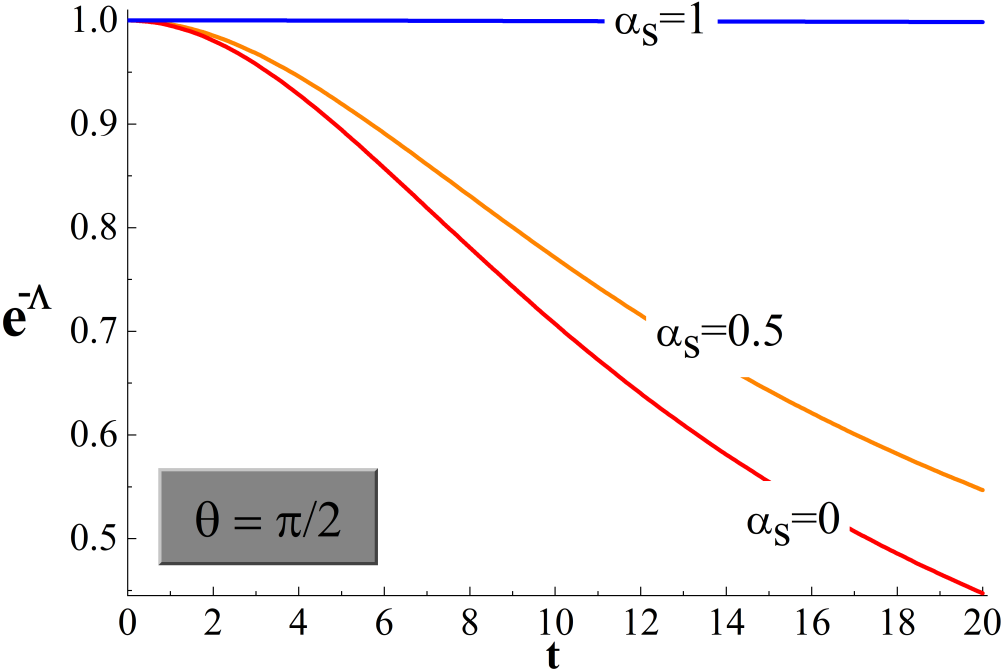}
\caption{(Color online) Time evolution of decoherence for different values of non-Hermiticity in the system, keeping the Hermiticity of the environment fixed, i.e., $\tau=0$.}
\label{fig3}
\end{figure}

Fig.\,\ref{fig3} shows the decoherence dynamics for different values of the system non-Hermiticity with the environment being Hermitian. Recall that the eigenvalues of the non-Hermitian system Hamiltonian are real as long as the parameter $\alpha_S$ satisfies the condition $|\alpha_S|\leq 1$. Therefore, the exceptional points correspond to the case when $\alpha_S=1$ as depicted by the blue line. It is interesting to observe a surprising result that the system being at the exceptional point shows a large coherence time regardless of the environment being Hermitian, at the exceptional point or non-Hermitian with $\mathcal{PT}$-symmetry, as also observed earlier both theoretically \cite{Chakraborty_Sarma_2019} and experimentally \cite{Wang_etal_2021}. But, we do see dependence of the system coherence on the environment when the system is not at the exceptional point, as observed in earlier figures. The system behaves as if it is at an exceptional point for $\theta = \frac{\pi}{2}$  when $\tau$ is large. That is when the Environment is highly non-Hermitian. The terms $\zeta^2 \cos^2(\theta)+\delta^2 \sin^2(\theta)$ and $2(\delta-\zeta) \cos(\theta) \sin(\theta)$ in \eqref{Decoherence_factor} give $\sin^2\theta + \cos^2\theta$ and $0$, respectively, for $\tau=0$, which means that there is no $\theta$ dependence in $\Lambda(t)$ when $\tau=0$. In order to get larger coherence times for the system, we can either tune $\alpha_S = 1$, bringing the system to the exceptional point, or we can tune $\theta = \frac{\pi}{2}$ for a non-Hermitian environment.
\section{Experimental schematic}\label{sec6}
With the advancement of $\mathcal{PT}$-symmetric and non-Hermitian physics, in particular, on the experimental front, preparation of non-Hermitian systems is not difficult. In general, there are two broad approaches to realize non-Hermitian dynamics in physical systems. One protocol involves embedding the system in a higher dimensional Hilbert space, for example, by coupling the system to an ancilla and then post-selecting the ancillary degrees of freedom to obtain effective non-Hermitian dynamics for the system. This method is familiar as the Naimark dilation \cite{Gunther_Samsonov_2008}. The other protocol relies on the Lindbladian formalism to study dissipation in the system and then obtaining the corresponding non-Hermitian Hamiltonian, see; for instance \cite{Minganti_Miranowicz_Chhajlany_Nori_2019}.

There exist numerous platforms for preparing non-Hermitian system Hamiltonians, such as, optical \cite{Guo_etal_2009,Ruter_etal_2010,Peng_etal_2014,Feng_El-Ganainy_Ge_2017}, topological \cite{Xiao_etal_2017}, acoustic \cite{Fleury_Sounas_Alu_2015,Shi_etal_2016}, N-V center \cite{Wu_etal_2019}, superconducting  \cite{Kawabata_Ashida_Katsura_Ueda_2018}, quantum dot \cite{Purkayastha_Kulkarni_Joglekar_2020}, etc. Indeed, several experimental studies have come up explicitly for the preparation of non-Hermitian spin-boson systems, see; for example \cite{Magazzu_etal_2018,Lambert_Ahmed_Cirio_Nori_2019}. A non-Hermitian two-level system can be obtained by embedding a transmon circuit in a resonator such that submanifolds comprising of the first and second excited states forms a two-level system. Under the assumption that the decay rate of the first excited state to the ground state as compared to the decay of the second excited state is much higher, it has been shown that the effective dynamics of the two-level system is governed by a non-Hermitian Hamiltonian as a result of the effect of the environment \cite{Naghiloo_Abbasi_Joglekar_Murch_2019}. 

$\mathcal{PT}$-symmetric non-Hermitian system have also been prepared using nitrogen-vacancy (NV) center in diamond \cite{Wu_etal_2019}. This approach utilizes the dilation method, extending a non-Hermitian $\mathcal{PT}$-symmetric Hamiltonian into a Hermitian Hamiltonian in a higher-dimensional Hilbert space. The NV center system consists of an electron spin ($S=1$) coupled to a $^{14}$N nuclear spin ($I=1$). The experiment is performed in a four-dimensional subspace spanned by $|m_s=0,m_I=1\rangle$, $|m_s=0,m_I=0\rangle$, $|m_s=-1,m_I=1\rangle$, and $|m_s=-1,m_I=0\rangle$. To implement the desired dilated Hamiltonian, two microwave pulses and two radio frequency wave pulses are applied to selectively drive the electron spin transitions and nuclear spin transitions, respectively. By engineering the amplitudes, frequencies, and phases of these microwave and radio-wave pulses, the desired Hamiltonian can be realized. For further details on this construction, one may refer to \cite{Wu_etal_2019}. 
\begin{figure}
\centering 
\includegraphics[scale=0.45]{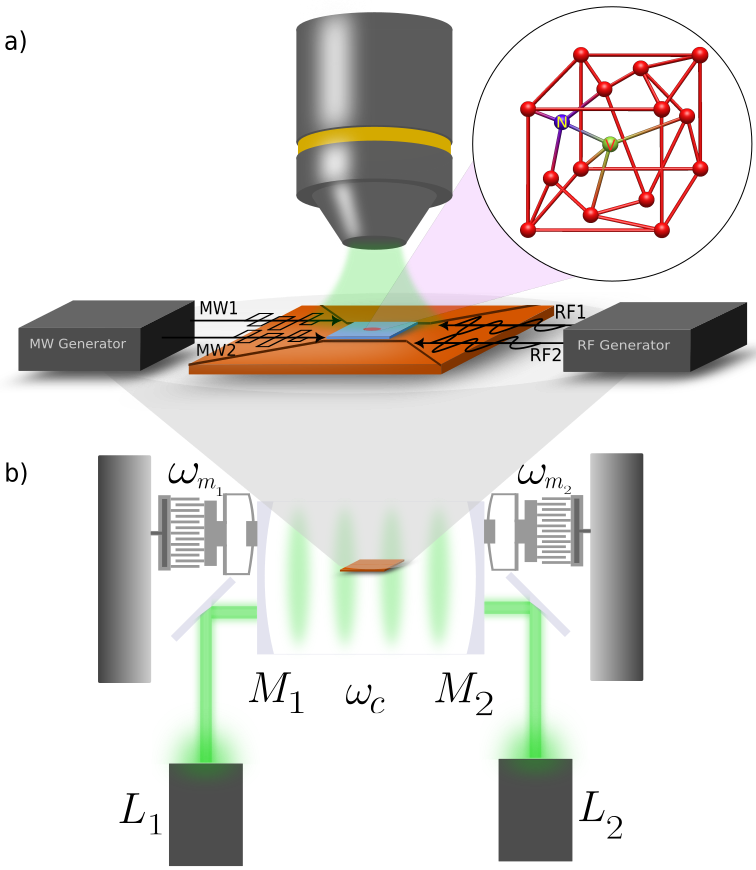}
\caption{(Color online) Schematic for the non-Hermitian system and non-Hermitian environment interacting model. Panel (a) shows the NV-center-based non-Hermitian system, whereas panel (b) provides the complete interacting model, where the NV-center-based non-Hermitian system is embedded within the optomechanical cavity non-Hermitian environment. The frequencies of the mechanical oscillators are $\omega_{m_1}$ and $\omega_{m_2}$, respectively. $\omega_{c}$ is the frequency of the cavity. $L_1$ and $L_2$ are the probe and driving laser, respectively.}
\label{experiment_diagram}
\end{figure}

While $\mathcal{PT}$-symmetric systems are well-realized, the understanding of the interaction of a $\mathcal{PT}$-symmetric systems with a $\mathcal{PT}$-symmetric environment has not been explored. In the earlier sections, albeit theoretically, we have studied it in detail. In the following, we prepare a detailed experimental schematic for the same that will propel experimental studies.

To realize a non-Hermitian $\mathcal{PT}$-symmetric environment, we propose employing optomechanical techniques as shown in Fig.\,\ref{experiment_diagram}. Specifically, we utilize a cavity with two mechanical oscillators positioned at opposite extremities. These two oscillators are fitted with two variable micro-electromechanical system (MEMS) oscillators that provide stable reference frequencies. MEMS oscillators transfer electrical energy into mechanical energy and are composed of a micro-mechanical resonator and a feedback amplifier to provide a sustainable resonance. MEMS oscillators provide many other advantages for such types of experiments; see, for instance, \cite{Feng_White_Hajimiri_Roukes_2008}. These oscillators, when coupled to the cavity, can potentially create a $\mathcal{PT}$-symmetric configuration by engineering the gain and loss rates to be balanced.

The NV center, embedded inside the cavity, serves as the system interacting with the non-Hermitian environment. While the precise implementation of the environment's Hamiltonian is beyond the scope of this work, prior studies have demonstrated the feasibility of using optomechanical setups to engineer non-Hermitian and $\mathcal{PT}$-symmetric dynamics \cite{Aspelmeyer_Kippenberg_Marquardt_2014, Jaramillo_etal_2020}. The schematic illustrates the conceptual experimental arrangement, leaving room for further exploration of the specific parameter requirements. Thus, the configuration can effectively generate a non-Hermitian setting, which may serve as an environment for any system. For the sake of convenience, we have chosen the system to be a non-Hermitian NV-center as depicted in panel (a). The panel (b) shows the complete non-Hermitian system and non-Hermitian interacting model.

All standard protocols for the optomechanical experiment \cite{Aspelmeyer_Kippenberg_Marquardt_2014} is to be followed. For example, the cooling mechanism to reduce the losses would be required. Experimental studies on $\mathcal{PT}$-symmetric system using open optomechanics treatment by taking the environmental losses of the optomechanics into account have come up; see, for instance, \cite{Jaramillo_etal_2020}, which can be used for a better understanding of the mechanism. Specific requirement for this study includes a proper adjustment of the detuning and other parameters, which will help in adjusting the non-Hermiticity of the environment. Also, the system and environment interaction angle can be controlled by changing the orientation of the NV-center spin.

\section{Conclusions and outlook}\label{sec7}
This work demonstrates the enhanced coherence times achievable by employing non-Hermitian $PT$-symmetric systems and environments. The study finds that maximal coherence preservation occurs when both the system and environment are non-Hermitian and $PT$-symmetric. Notably, the results challenge the prevailing view that stronger environmental coupling invariably hastens decoherence. Experimental schematics are proposed, employing NV centers and optomechanical systems to realize and investigate these theoretical insights. These advancements pave the way for innovative methods to safeguard quantum information, addressing a critical bottleneck in quantum computing development.

The exploration of non-Hermitian $PT$-symmetric systems opens new frontiers in quantum information science. Future research may focus on refining the experimental realization of such systems, especially in diverse platforms such as photonics, superconducting circuits, and NV centers. Additionally, investigating the interplay of $PT$-symmetric systems with different types of environments, such as spin environments or engineered reservoirs, could provide deeper insights into decoherence control. Extending this framework to multi-qubit systems and complex quantum networks is another promising direction. These advancements could contribute significantly to the development of robust quantum technologies, overcoming the decoherence challenges in scalable quantum computing and communication. \\

\noindent \textbf{\large{Acknowledgments:}} Duttatreya is supported by a UGC-NET PhD Research Scholarship. S.D. acknowledges the support of research grants DST/FFT/NQM/QSM/2024/3 (An initiative under the National Quantum Mission of DST, Govt. of India), SB/SRS/2022-23/89/PS (by DST-SERB, Govt. of India), and NFSG/GOA/2023/G0928 (by BITS-Pilani).



\end{document}